\documentclass[webconf, onecolumn]{svjour}
\usepackage{graphics}
\usepackage[varg]{txfonts} 
\usepackage[latin1]{inputenc}
\session-title{AO for ELT, June 2009}
\begin{document}
\title{Characterisation of the influence function non-additivities for a 1024-actuator MEMS deformable mirror}
\author{C\'elia Blain\inst{1}\fnmsep\thanks{\email{cblain@me.uvic.ca}} \and Rodolphe Conan\inst{1} \and Colin Bradley\inst{1} \and Olivier Guyon\inst{2} \and Curtis Vogel\inst{3}}
\institute{University of Victoria, Department of Mechanical Engineering, PO Box 3055, Stn. CSC, Victoria, BC, V8W 3P6, Canada\and Subaru Telescope NAOJ, 650 N. A'Ohoku Place, 96720 Hilo, HI, USA \and University of Montana, Department of Mathematical Sciences, Bozeman, MT 59717-2400, USA}

\abstract{
In order to evaluate the potential of MEMS deformable mirrors for open-loop applications, a complete calibration process was performed on a 1024-actuator mirror. The mirror must be perfectly calibrated to obtain deterministic membrane deflection. The actuator's stroke-voltage relationship and the effect of the non-additivity of the influence functions are studied and finally integrated in an open-loop control process. This experiment aimed at minimizing the residual error obtained in open-loop control.
} 
\titlerunning{MEMS non-additivity characterisation}
\maketitle
Proceedings of the first AO for ELT conference, June 2009, Paris. 
Copyright ownership: EDP Sciences, the original publication is available at www.edpsciences.org

\section{Introduction}
\label{sec:intro}
In the prospective of new Extremely Large Telescopes (ELTs) and their related science instruments, the need for open-loop (OL) control of deformable mirrors (DM) is increasing. Much of the research in this area focuses on the development of an accurate model\cite{D.Gavel_08}\cite{C.Blain_08}\cite{K.Morzinski_07}\cite{J.B.Stewart_07}\cite{C.R.Vogel_06} for Micro-Electro-Mechanical-Systems (MEMS) DMs. 
With a continuous membrane DM, each actuator deflection is the direct result of the voltage it receives combined with the indirect effect of the vertical deflection of the actuators surrounding it.
In order to build an accurate model of a MEMS DM with a continuous membrane, one needs to quantify both the non-linear relationship between the input voltage and the resulting actuator deflection as well as the effect of the coupling (through the membrane) between neighbouring actuators.
In this paper, we first present the results of a thorough DM characterisation. Sec.~\ref{sec:lin} is focused on the stroke-voltage relationship  while Sec.~\ref{sec:charaIFnl}  describes the influence function (IF) non-additivity. Finally, Sec.~\ref{sec:ol}, illustrates the benefit that such characterisation can provide for 'model free' OL control of MEMS.  
\section{Experimental setup}
\label{sec:exp_setup}
The experimental setup consists of a 1024 actuators Boston Micromachines MEMS DM with 150 Volts (14 bit resolution) electronics manufactured by NASA JPL. A Zygo PTI 250 interferometer is positioned in front of the DM. The interferometer beam pass through a density filter to improve the fringe contrast. A first computer is dedicated to the Zygo interface (metrology software and interferometer command) while a second computer controls the DM electronics, initiates the interferometer measurement and transfers the recorded data to the laboratory data server. DM and interferometer are set on a vibration isolation optical table.

\section{Calibration of the actuator stroke-voltage relationship}
\label{sec:lin}

For MEMS, the stroke-voltage relationship for an actuator $i$ is quadratic and can be described by the following equation:
 \begin{center}
	\begin{equation}
	   stroke(i) = gain(i).\:V(i)^{2}\:+\: bias(i)
	   \label{strVSvolt}
	\end{equation}
  \end{center}
where \textit{V} is the voltage sent to the actuator, \textit{bias} is the actuator offset and \textit{gain} is the actuator gain.

\begin{figure}
\centering
\resizebox{0.497\columnwidth}{!}{\includegraphics{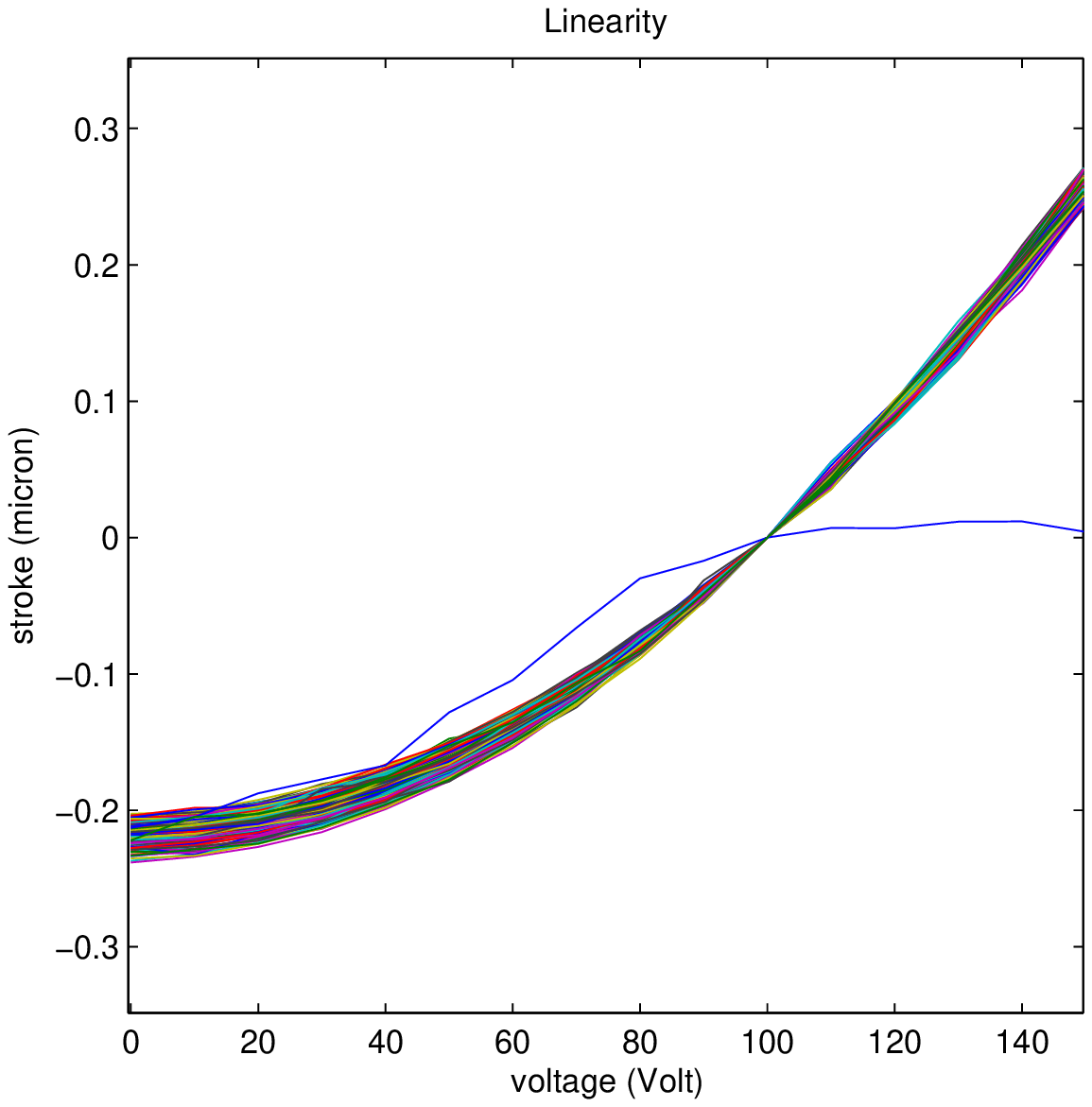} } 
\resizebox{0.497\columnwidth}{!}{\includegraphics{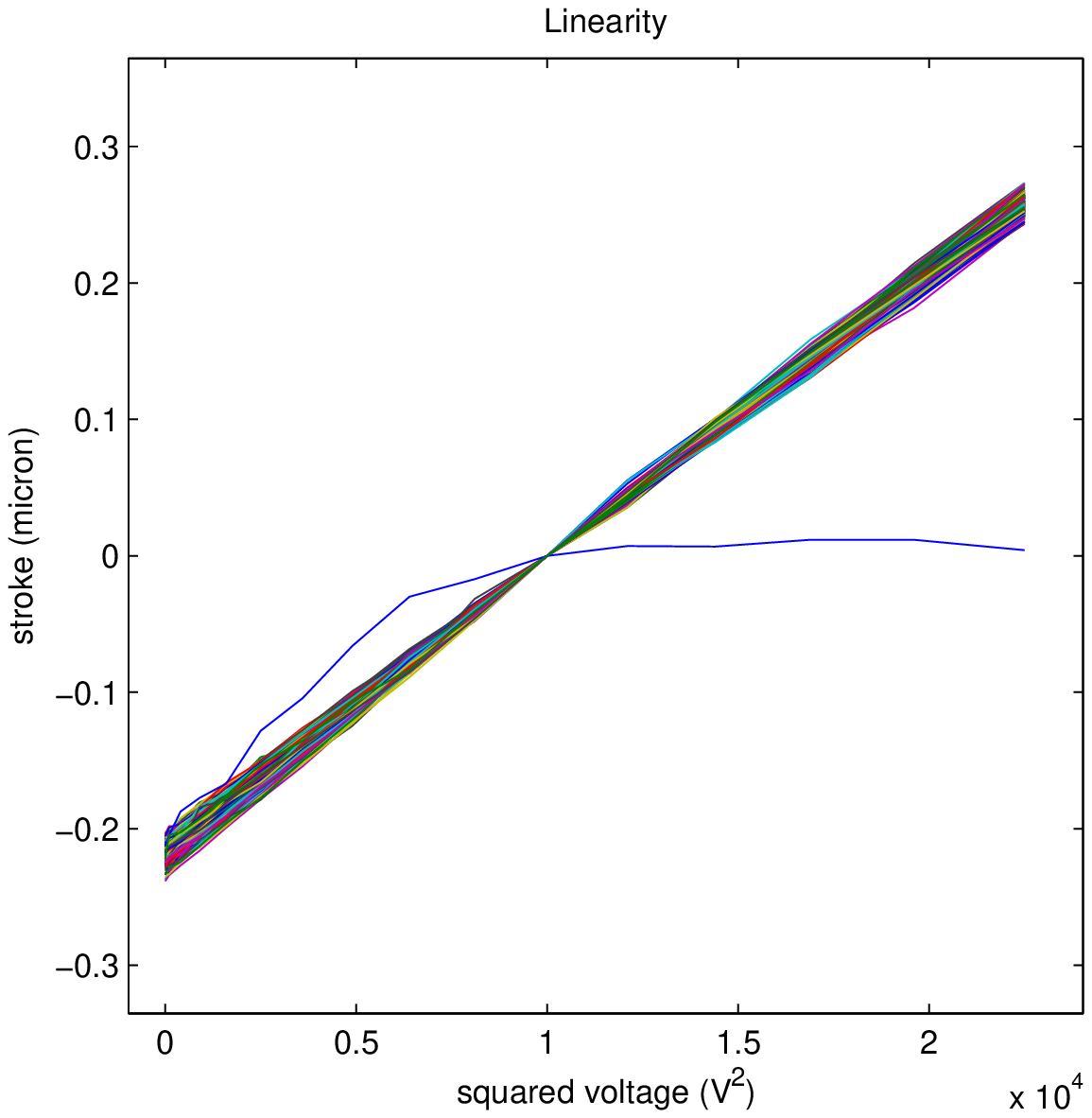} }
\begin{center}  (a)    \hspace{ 7 cm}   (b) \end{center}
\caption{(a) Stroke-voltage relationship and (b) stroke-squared voltage relationship, for the 324 actuators. All actuators present a maximum stroke of approximately 0.5 micron. The blue line corresponds to a defective actuator, coupled with an actuator located outside the array of interest. }
\label{fig:lin}      
\end{figure}

The precise calibration of each actuator's stroke-voltage relationship is a critical step toward accurate open-loop control of the deformable mirror. For this experiment, due to a large number of malfunctioning actuators located on the right side of the DM, an array of 18 by 18 actuators (total of 324 actuators) is selected. Only one actuator among the 324 is coupled with an actuator located outside of the array. This actuator presents a reduced maximum stroke and can be seen in Fig.~\ref{fig:lin} (blue curve). With a maximum voltage output of 150 Volts, the resulting maximum stroke is approximately 0.5 micron. In Sec.~\ref{sec:ol}, Eq.~\ref{strVSvolt} will be inverted to obtain the input voltage from the desired stroke and inject it in the open-loop control process.

\section{Characterization of influence function non-additivity}
\label{sec:charaIFnl}

\begin{figure}
\centering
\resizebox{0.497\columnwidth}{!}{\includegraphics{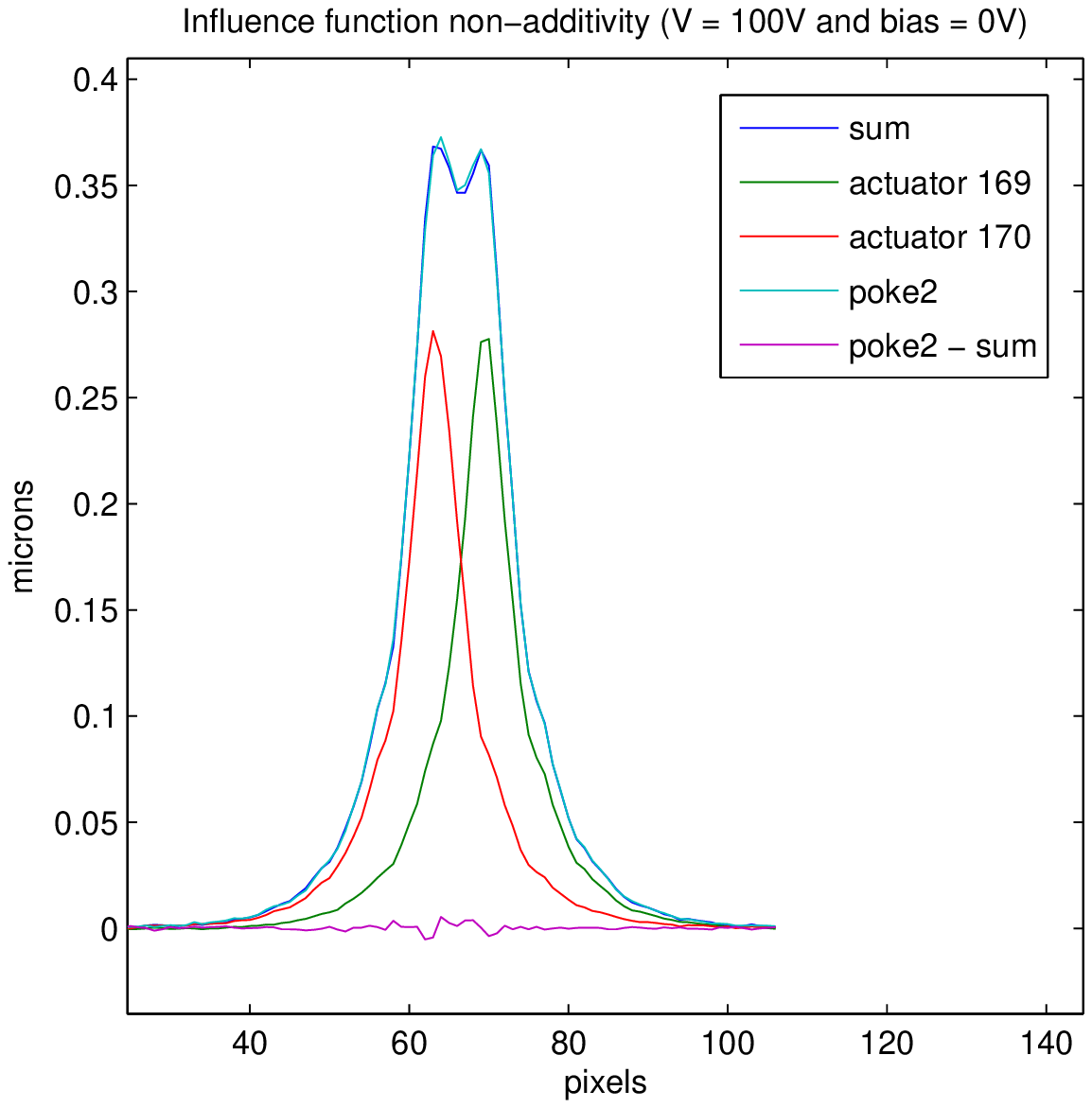} } 
\resizebox{0.497\columnwidth}{!}{\includegraphics{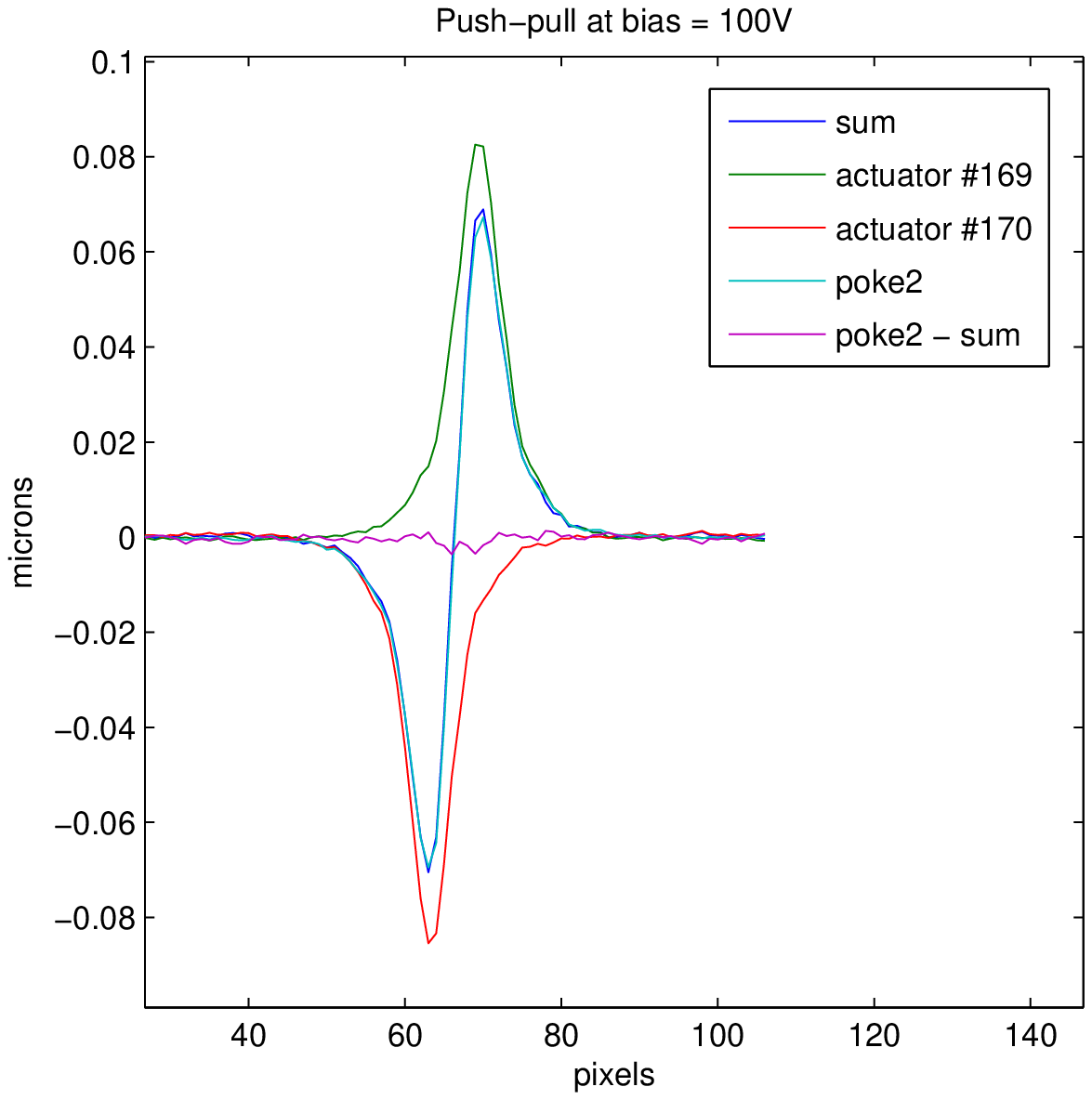}  }
\caption{(a) Influence function non-additivity measurement for a pair of actuators and (b) push-pull configuration.}
\label{fig:NAdemo}      
\end{figure}

To study the influence function non-additivity, several random pairs of neighbouring actuators within the array of interest are selected. This section will present only the results obtained for the pair of actuators indexed 169 and 170, which are representative of the other actuators tested. Three tests were conducted to characterise the non-additivity.

First, for each pair of actuators, the membrane was set to a bias voltage of 0 Volt. The 2 actuators under test were first push independently to 50 Volts, 100 Volts, and 150 Volts. Then, the 2 actuators were pushed simultaneously to these 3 voltages.  

This test is repeated with a bias voltage of 150 volts while the actuators are released independently then simultaneously to 0 Volt, 50 Volts, and 100 Volts. 

Finally, a third test focus on the non-additivity effects when the pair of actuators are on a push-pull configuration.\\

Fig.~\ref{fig:NAdemo} (a) shows the 'contour cut' of the influence function when each actuator is pushed independently (green and red contours) then simultaneously (light blue contour labeled \textit{poke2}) to 100 Volts while the bias voltage is 0 Volts. The linear sum (obtained numerically after the measurement) is also presented (dark blue contour, labeled \textit{sum}) as well as the difference between the poke2 and the linear sum (purple plot, labeled \textit{poke2 - sum}).

\subsection{Pair of neighbouring actuators}
\label{sec:2x2}

\begin{figure}
\centering
\resizebox{0.497\columnwidth}{!}{\includegraphics{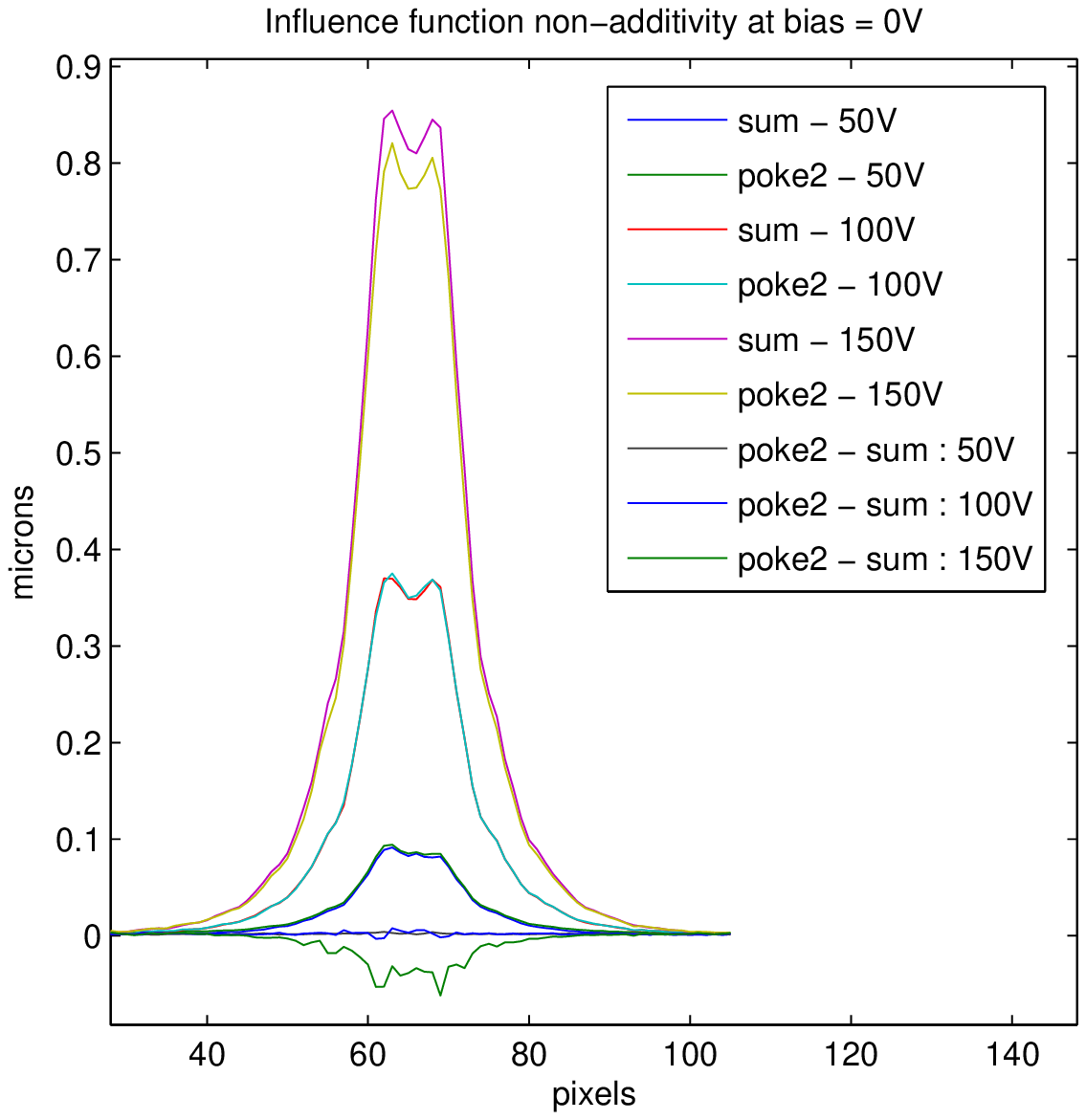}  } 
\resizebox{0.497\columnwidth}{!}{\includegraphics{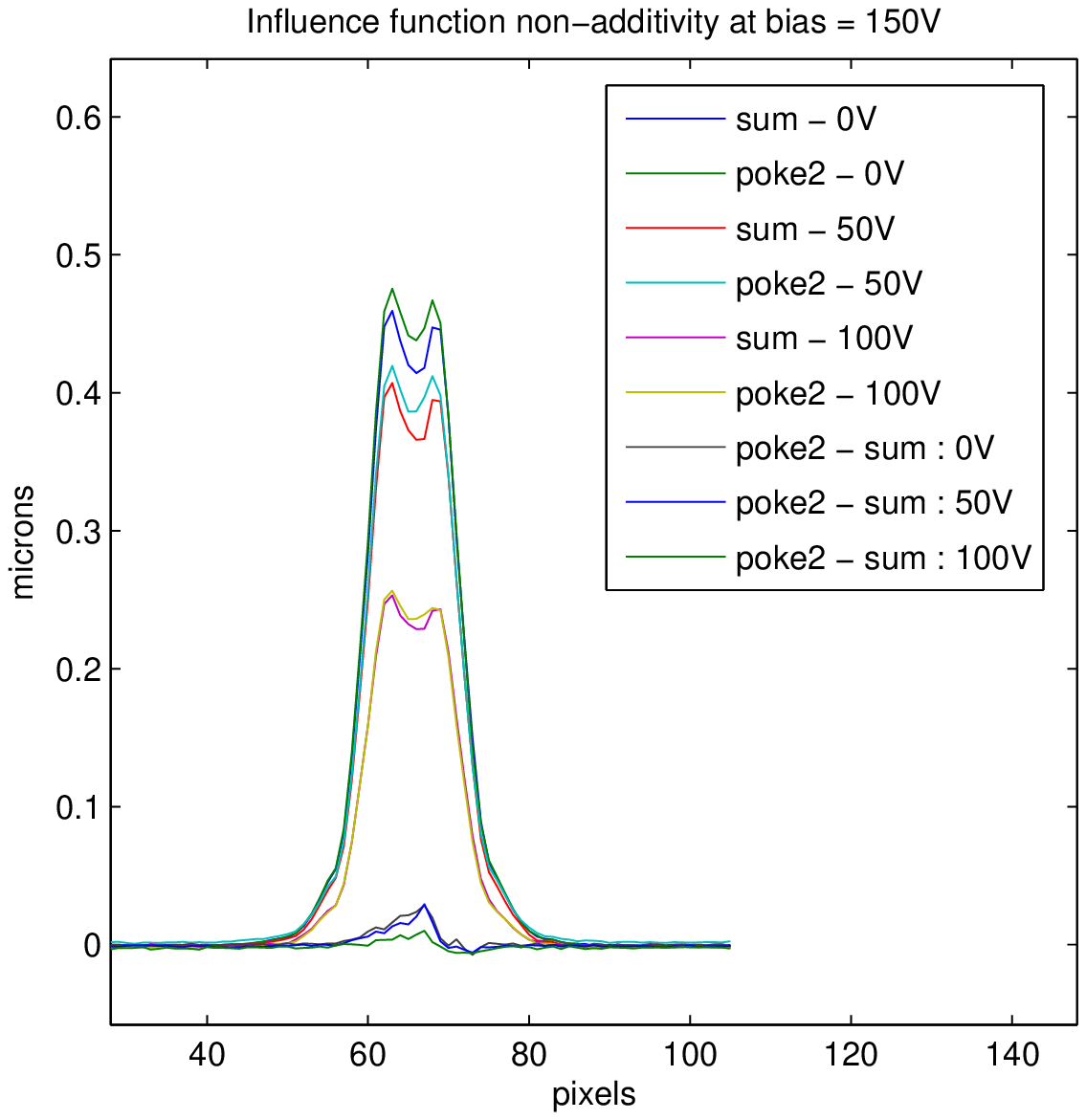}  }
\begin{center}  (a)    \hspace{ 7 cm}   (b)  \end{center}
\caption{Non- additivity of influence function measurement for a pair of actuators with (a) a bias of 0 Volt and (b) a bias of 150 Volts.}
\label{fig:NA2-2}      
\end{figure}

The results obtained with the pair of neighbouring actuators indexed 169 and 170 are presented in Fig.~\ref{fig:NA2-2}. 

(a) corresponds to a bias of 0 Volt while (b) corresponds to a bias of 150 Volts.
From here forward, the \textit{relative voltage} will be defined as the difference between the bias voltage (applied to the whole DM) and the input voltage (only applied to the pair of  neighbouring actuators under test).
For example, for a bias voltage of 150 Volts, and an input voltage of 100 Volts, the relative voltage is 50 Volts or for a bias voltage of 0 Volt, and an input voltage of 50 Volts, the relative voltage is also 50 Volts. Fig.~\ref{fig:NA2-2} reveals four remarkable non-linear behaviours.
\begin{itemize}
\item The strokes obtained with a 0 Volt bias are much larger than the strokes obtained with a 150 Volts bias for the entire range of input voltage values. 
\item The linear sum (labeled \textit{sum}) is always providing a smaller stroke than the pair of actuators poked simultaneously (labeled \textit{poke2}) except in one particular case, when the bias voltage is 0 Volt and the input voltage is 150 Volts.
\item  The difference in stroke between a relative voltage of 100 Volts and a relative voltage of 150 Volts, varies from approximately 0.5 microns with a bias of 0 Volt to less than 0.1 microns with a bias of 150 Volts.
\item Unlike preceding models \cite{C.R.Vogel_06} of 2 actuators poked simultaneously and presenting a resulting influence function peak shaped with a flat top, this experiment reveals the presence of a dip at the top of the peak.
\end{itemize} 
To explain these behaviours, several factors need to be taken into account. First, at 0 Volt bias (corresponding to the default DM position), the membrane is relatively flat and undergoes no stretch. However, at 150 Volts bias, the membrane becomes slightly dome shaped and stretched due to the fact that the edge actuators have less available stroke than the actuators located closer to the center. This additional stretch decreases the elasticity available for the pair of actuators under test, resulting in a smaller stroke.  
This elasticity saturation can also explain the difference in stroke between similar relative voltages (100V and 150V) for the 2 different bias (0V and 150 V). 

Another factor that contributes to this difference relies on the stroke-voltage relationship of each actuator. Fig.~\ref{fig:lin} (a), presents the typical quadratic relationship between stroke and voltage. At the beginning, the stroke increase is much slower than the voltage increase. After a midpoint, located around 100 Volts for this apparatus, the stroke increase is much faster and follows the voltage increase in a near linear fashion. A relative voltage of 50 Volts in the 0 V bias configuration corresponds to a slow stroke increase thus providing a small stroke. This same relative voltage in the 150 Volts bias configuration corresponds to a point further up the curve, providing a larger stroke.

\subsection{Push-pull}
\label{sec:pushpull}

In a push-pull configuration, the pair of actuators are pushed in opposite direction.  As opposed to other DM technology where the actuation is 'bidimentional', up or down (such as piezo-stack DM), the MEMS DM presents a characteristic pull-in behaviour. In order to set the actuators in a push-pull configuration, we added a bias of 100 Volts then set the actuator indexed 169 to 77 Volts and the actuator 170 to 118 Volts. This provides a peak to valley stroke of 0.2 micron. This stroke is kept relatively small to limit the stress undergoes by the membrane and avoid to damage it. 

Fig.~\ref{fig:NAdemo} (b) shows that the non-additivity of influence function is compensated when the actuators are in this configuration (plots \textit{sum} and \textit{poke2} are on top of each other). This is an encouraging result since the actuators configuration when the DM try to compensate for atmosphere turbulence is close to a push-pull.

\subsection{Array of 3 by 3}
\label{sec:3x3}
\begin{figure}
\centering
\resizebox{0.497\columnwidth}{!}{\includegraphics{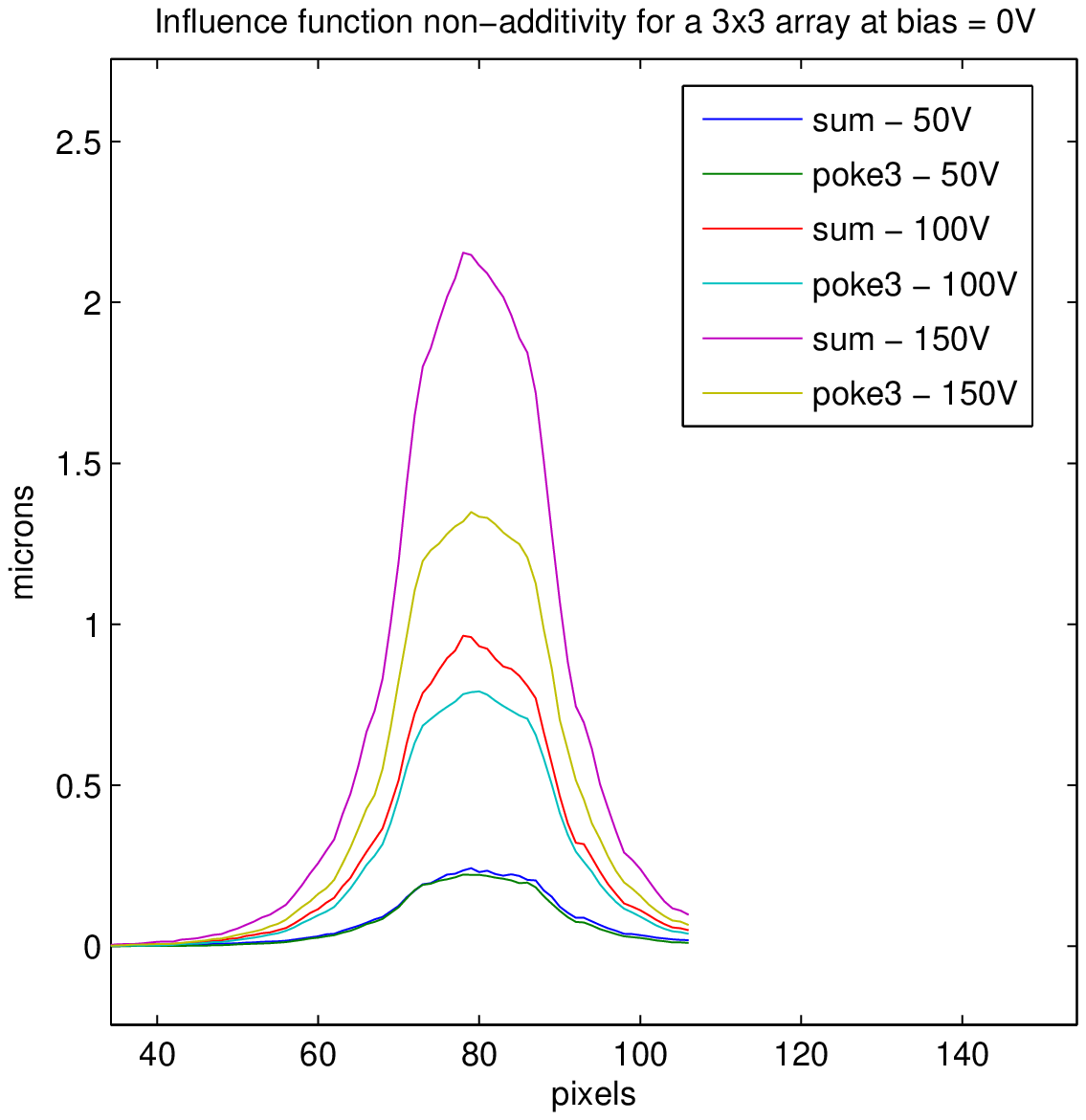}  } 
\resizebox{0.497\columnwidth}{!}{\includegraphics{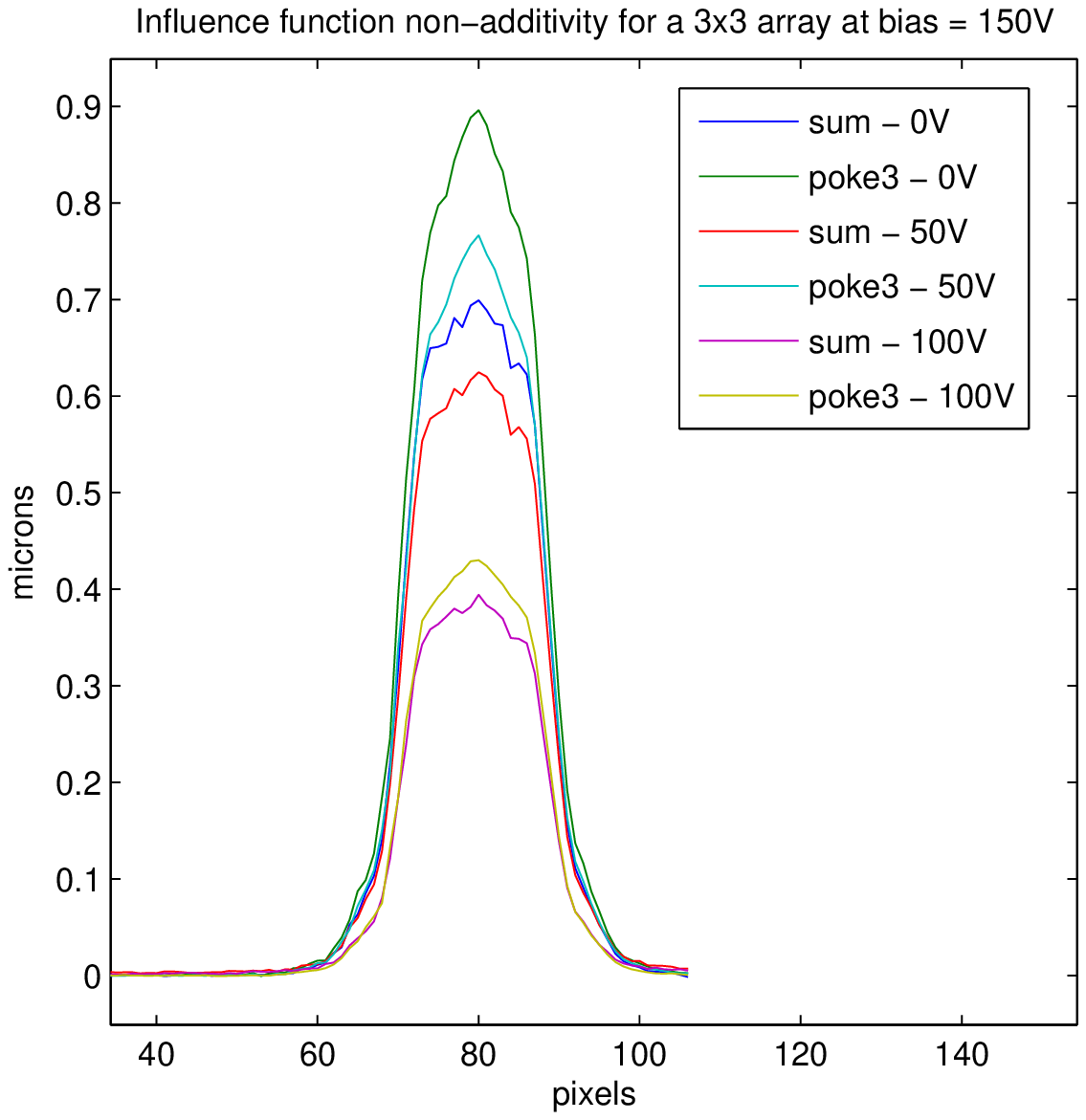}  }
\begin{center}  a)    \hspace{ 7 cm}   b)  \end{center}
\caption{Non- additivity of influence function measurement for an array of 3 by 3 actuators with (a) a bias of 0 Volt and (b) a bias of 150 volts.}
\label{fig:NA3-3}      
\end{figure}

The test described in Sec.~\ref{sec:2x2} are repeated with an array of 3 by 3 actuators.
Fig.~\ref{fig:NA3-3} (a), presents the comparison between the 3 by 3 actuators which are first pushed together (labelled \textit{poke3}) and second pushed independently and added numerically (labelled \textit{sum}) at relative voltages of 50 Volts, 100 Volts and 150 Volts. Fig.~\ref{fig:NA3-3} (b) presents these results when the bias is 150 Volts.
At 0 Volt bias, the linear sum always provides a larger stroke than pushing the actuators simultaneously. However, at 150 Volts bias, the opposite effect occurs and the \textit{poke3} is always larger than the linear \textit{sum}. An array of 3 by 3 actuators pushed together has a much larger strength than each actuator pushed independently . Thus, when the membrane elasticity is reduced due to stretching, the \textit{poke3} gives larger stroke than the linear sum.
The stretching effect is also responsible for the large variation between the stroke obtained for a relative voltage of 150 Volts at 0 Volts bias (approximately 1.4 microns) and at 150 Volts bias (approximately 0.9 microns). 
Finally, at 0 Volt bias, the influence function presents a much larger profile than the one obtained at 150 Volts bias. 

\section{Open-loop control performance}
\label{sec:ol}

The goal of this section is the DM open-loop control using only the calibration of the actuator's stroke-voltage relationship and influence function, studied in Sec.~\ref{sec:lin} and Sec.~\ref{sec:charaIFnl}. 

First, 100 computer generated phase screens are simulated to match the atmosphere turbulence as seen by a 30 m diameter telescope ($r_{0}$ = 15 cm, $L_{0}$ = 60 m, wind speed(unirectionnal) = 10 m.$s^{-1}$) then scaled to match the DM's maximum stroke. The least squared fit of the phase screens projected onto the normalised influence functions provides the actuator stroke maps.
Fig.~\ref{fig:lin} (b), shows that the stroke and the squared voltage follow a linear pattern. The \textit{gain} and \textit{bias} coefficients needed in Eq.~\ref{strVSvolt} are extracted from this stroke-voltage characterisation. The voltage maps to be sent to the DM are finallly deduced from the stroke maps by reversing Eq.~\ref{strVSvolt}. 

\begin{figure}
\centering
\resizebox{0.75\columnwidth}{!}{\includegraphics{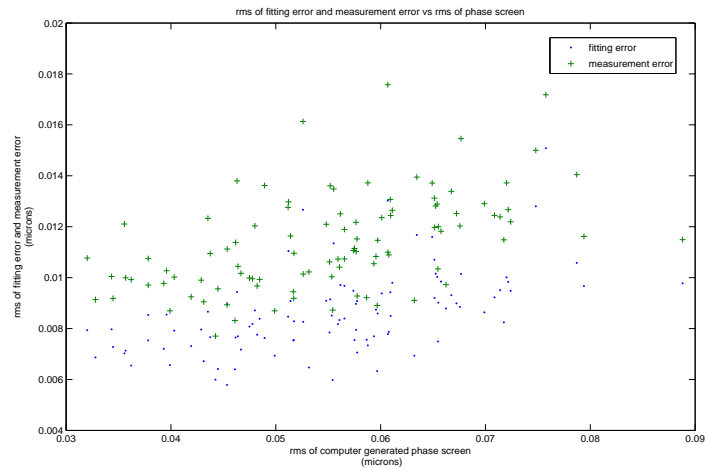} }
\caption{Open-loop rms and fitting error rms versus phase screen rms.}
\label{fig:olresults}      
\end{figure}

The multiplication of the stroke maps by the normalised influence functions gives the fitted phase screens. The fitting error is the result of the difference between the original phase screens and the fitted phase screens. The fitting error corresponds to the DM sampling error due to the limited number of actuators.
Fig.~\ref{fig:olresults} gives the rms of the fitting error and the rms of the measurement error (or open-loop error) as a function of the rms of the generated phase screen.

A subsequent paper (Blain et al., 2009) is dedicated to the optimisation of MEMS open-loop control using only the characterisation of the actuator's stroke-voltage relationship and influence function. This paper presents the open-loop performance achieved for a 1024-actuator MEMS driven with an upgraded electronic (maximum output voltage = 200 Volts). Tab.~\ref{tab:1}  gives an overview of the mean rms results obtained with this upgraded apparatus.  The DM error, corresponding to the difference between the open-loop error and the fitting error, gives an estimate of the error due to the DM non-linear effects.
\section{Conclusion}
\label{sec:concl}
\begin{table}
\centering
\caption{Presentation of the mean rms values for 100 phase screens.}
\label{tab:1}       
\begin{tabular}{lllllll}
\hline\noalign{\smallskip}
phase scr. & proj. on IF & fitting error & open-loop error & DM error & $ \frac{open\,loop}{phase\,scr.} $ & $ \frac{DM}{phase\,scr.} $ \\
\noalign{\smallskip}\hline\noalign{\smallskip}
96.73nm & 95.78nm  & 13.30nm & 16.52nm & 10.78nm & 17.28 $\%$ & 11.23 $\%$\\ 
\hline\noalign{\smallskip}
\end{tabular}
\end{table}
In this paper, the influence function non-additivity are linked to the combination of elasticity loss (due to membrane stretching) and to the actuator's quadratic stroke-voltage relationship. It has been shown that these IF non-additive effects are cancelled in a push-pull configuration. As a result, the IF non-additivity may prove to be negligible when a random shape (such as a turbulent phase screen) is applied to the DM.
The characterisation of the actuator stroke-voltage relationship and the actuator influence function are integrated in order to control the DM in open-loop.
Encouraging performances are obtained. With original mean phase screens of 96.73 nm rms,  the mean open-loop rms obtained is 16.52 nm and the mean fitting error rms is 13.30 nm. A more detailed study of the open-loop control of MEMS through actuator calibration can be found in a subsequent paper (Blain et al., 2009).


\end{document}